\documentclass[a4paper,12pt]{article}
\usepackage{graphicx} 
\usepackage{amsmath}
\usepackage{amssymb}
\usepackage{epsfig}
\usepackage{xcolor}
\usepackage{hyperref}
\usepackage{cite}
\usepackage{graphicx}
\usepackage{subfigure}
\usepackage[normalem]{ulem}
\newcommand{\nc}{\newcommand}
\nc{\beq}{\begin{equation}} \nc{\eeq}{\end{equation}}
\nc{\beqa}{\begin{eqnarray}} \nc{\eeqa}{\end{eqnarray}}
\nc{\ba}{\begin{array}} \nc{\ea}{\end{array}}

\def\al{\alpha}
\def\be{\beta}

\def\la{\lambda}

\def\pa{\partial}

\def\ph{\varphi}

\def\Ga{\Gamma}

\begin{document}
\begin{center}
{\bf \Large All-loop effective potential for arbitrary  \\[0.2cm] scalar models in curved space-time} \vspace{1.0cm}

{\bf \large  V.A. Filippov$^{1,a}$, R.M. Iakhibbaev$^{1,b}$  \\[0.3cm] and D. M. Tolkachev$^{1,2,c}$  }

\vspace{0.5cm}
{\it $^1$Bogoliubov Laboratory of Theoretical Physics, Joint Institute for Nuclear Research, 6, Joliot Curie, 141980 Dubna, Russia\\ and \\
$^2$ Stepanov Institute of Physics,
68, Nezavisimosti Ave., 220072, Minsk, Belarus\\}
\vspace{0.5cm}

\abstract{In this paper we calculate the leading divergences of the effective potential for an arbitrary scalar theory on a curved spacetime background. Based on the recurrence relation between the leading poles following from the locality condition, we obtain a system of generalised renormalisation-group equations that can be studied numerically or analytically in some special cases. We study the simplest effective potentials for power-like models and give a comparison of them in the framework of cosmological phenomenology.}
\\
\textit{E-mails}: {$^a$vafilippov@theor.jinr.ru, $^b$yaxibbaev@jinr.ru, $^c$den3.1415@gmail.com }
\end{center}

\section*{Introduction}

\; \; The effective potential plays an important role in the study of the ground state structure in quantum field theory \cite{Coleman:1973jx,Jackiw:1974cv}. Calculations of radiative corrections in the study of vacuum stability for scalar theories were carried out in both flat space \cite{Iliopoulos:1974ur,McKeon:2011vs,Kastening:1991gv}, and curved space \cite{Ishikawa:1983kz,Buchbinder:1983vve,Buchbinder:1984vys,Bunch:1981tr,Ford:1981xj,Toms:1982af}. The Renormalisation group (RG) method is of great importance in these studies, because since it allows one to talk about the behavior of the perturbation theory series for the effective potential in different approximations and in different regimes. However, the study of radiative corrections for scalar theories on a curved background in the framework of general relativity theory or its extensions.  is limited by the renormalisability condition of the classical interaction \cite{Barvinsky:1993zg}, although the majority of most interesting and reliable cosmological potentials are non-renormalisable  from the point of view of quantum field theory (see., e.g, \cite{Martin:2013tda}). 

As is known, the main obstacles for the perturbative analysis of non-renormalisable quantum field models are the unpredictable structure of counter-terms and the dependence of the results of loop calculations on the subtraction scheme, which makes it impossible to carry out the renormalisation procedure in the sense of the usual redefinition of the parameters of the initial model. Nevertheless, there are a number of papers generalising the standard RG approach \cite{Bork:2015zaa,Kazakov:2022pkc,Kazakov:2023tii,Kazakov:2020kbj}. The apparatus developed in these works makes it possible to sum up all contributions within the framework of perturbation theory. However, only the leading order (leading poles or logarithms) are scheme-independent. In this work, we apply the approach of the generalised renormalisation group developed in the aforementioned papers to the study of scalar field theory with arbitrary interaction $V_0(\varphi)$ on a curved spacetime background with non-minimal conformal coupling to gravity\footnote{In this work we put $m_P=(8\pi G_N)^{-1/2}=1$ where $G_N$ is the gravitational constant.}:
\beq
S[g_{\mu \nu},\ph]=\int d^4x \sqrt{-g} \left(\dfrac{1}{2} R + \dfrac{1}{2}g_{\mu \nu}\partial^\mu\varphi\partial^\nu\varphi + \dfrac{1}{2} \xi R \varphi^2 - \la V_0(\varphi)\right) \label{action},
\eeq
where $g_{\mu \nu}$ is the metric tensor, $\la$ is the coupling constant, $R$ is the Ricci scalar, and $\xi R \varphi^2$ is the non-minimal scalar-tensor coupling term, providing covariance of the scalar field equations under conformal transformations of the metric, so $\xi$ is the non-minimal coupling constant \cite{Birrell:1982ix}. Special cases of this kind of model are often used in studies of the cosmological inflation models, e.g. \cite{Bezrukov:2010jz, Inagaki:2014wva, Kallosh:2013pby,Chiba:2014sva}. 

We give calculations of a number of vacuum Feynman diagrams in the first orders of perturbation theory for the effective potential in the approximation of linearity in curvature $R$ on the basis of the local-momentum representation formulated using Riemannian normal coordinates. Using the Bogoliubov-Parasyuk theorem \cite{BP,Hepp,Zimmermann} on the locality of counter terms, one can relate the leading divergences in these diagrams by a recurrence relation and obtain a generalised RG-equation. The obtained RG-equation has the known analytical solution in the limit of the renormalisable case of interactions, but for higher degree potentials (non-renormalisable case) we give only a qualitative numerical analysis of the behavior of the effective potential, since the obtained equation is essentially nonlinear, which is a standard feature of RG-equations for non-renormalisable models. Finally, we give numerical estimates of the consequences of the application of the investigated effective potentials for inflationary cosmology.

\section{Effective potential on curved spavetime background}

The effective potential is part of the effective action that does not depend on the momenta (or derivatives), and the effective action itself is represented as an expansion \cite{Buchbinder:2017lnd}
\beq
\Ga[\ph,\,g_{\mu\nu}]\,=\,
\int d^4x\,\sqrt{-g}\Big\{
-\,V_{eff}(\ph) + \frac12\,Z(\ph)
\,g_{\mu\nu}\,\pa^\mu\ph\,\pa^\nu\ph
+ ... \,\Big\}\,,
\label{EffPot}
\eeq
and is the result of the Legendre transform of the generating functional, where $V_{eff}(\ph)$ is the effective potential, $Z(\ph)$ is a correction to the wavefunction.
A direct way of perturbative calculation of the effective potential is to sup up all vacuum one-particle-irreducible diagrams obtained by Feynman rules derived from the shifted action $S[\ph+\hat{\ph}]$ where $\ph$ is a classic field, obeying a classical equation of motion, so $\hat{\ph}(x)$ is a quantum field \cite{Jackiw:1974cv,Iliopoulos:1974ur}. The effective potential can be expanded by the coupling constant:
\beq
V_{eff}=\lambda \sum_{k=0}^\infty \left(- \dfrac{\lambda}{16\pi^2}\right)^k V_k,
\eeq
where $V_k$ is $k$-th order correction of the potential.

However, whereas Feynman's rules for computing corrections in flat space are derived simply from the direct action decomposition, for the computation of corrections in curved space one has to use more sophisticated techniques, e.g., local-momentum representation \cite{Bunch:1979uk,Buchbinder:2017lnd}. In this paper, we prefer the latter representation. Its main advantage is that all calculations can be performed in flat spacetime (but using modified Feynman rules for propagators and vertices). So the result for some local quantity can always be represented in a covariant form.  In order to obtain the modified Feynman rules, it is necessary to perform the expansion  \cite{Bunch:1979uk}:
\beq
g_{\al\be}(x) = g_{\al\be}(x^{\prime})
-\frac13\,R_{\al\mu\be\nu}(x^\prime)\,y^\mu \,y^\nu\ + \ldots
\label{expansion 2}
\eeq
in deviation $y^\mu=x^\mu-x'^\mu$ where the coordinates of any fixed point $x'^\mu$ (at this point the manifold can be defined by $g_{\mu\nu}=\eta_{\mu\nu}$, i.e. manifold is locally flat) and $x^\mu$ is in neighborhood of $x'^\mu$ on the Riemannian manifold. In \eqref{expansion 2} the dots imply the omitted terms of higher orders by curvature. From this representation, by means of the equation
\beq
H(x,x') G(x,x')=-\delta(x-x'),
\eeq
where $H(x,x')$ is the bilocal operator for the expanded action \eqref{action} (its explicit form is given in \cite{Bunch:1979uk,Sobreira:2011ep}) it is possible  to extract a modified Feynman propagator  
\beq
G(y)\,=\,\int \frac{d^4k}{(2\pi)^4}\,e^{iky}\,
\Big\{\,
\frac{1}{k^2+{\bar{m}}^2}
\,-\,\frac{(\xi-1/6)\, R}{(k^2+\bar{m}^2)^2}+ \ldots
\Big\}\,
\label{propagatorR}
\eeq
where the effective mass is $\bar{m}^2=\la V_0''(\ph)$ (further, for convenience, the derivatives from the potential are denoted as $V^{(n)}_0(\ph)=v_n$) and the dots again refer to higher curvature corrections. 

Thus, we can construct the first loop diagrams and compute them in the standard way as $-\text{tr} \log G(x,x')$ (see e.g., \cite{Bunch:1979uk,Sobreira:2011ep}). In the dimensional regularisation $d=4-2\epsilon$ we have
\beq
V_1=\textbf{V}_1+\textbf{W}_1
\eeq
that explicitly has the following form:
\beq
\textbf{V}_1=\frac{1}{4}v_2^2\left(\frac{1}{\epsilon}+\log\left(\frac{\mu^2}{\bar{m}^2}\right)\right)
\label{1loopflat},
\eeq
\beq
\textbf{W}_1=\frac{1}{2}v_2\hat{\xi} R\left(\frac{1}{\epsilon}+\log\left(\frac{\mu^2}{\bar{m}^2}\right)\right),
\label{1loopcurved}
\eeq
here $d$ is the number of dimensions, $\mu$ is the dimensional transmutation parameter, $v_2=V_0''(\ph)$, and $\hat{\xi} =1/6-\xi$, and the result is given here only for poles and logarithms, i.e. constants are omitted. These expressions correspond to calculations known in the literature \cite{Sobreira:2011ep,Ford:1981xj}. Here $\textbf{V}_1$ is the term corresponding to the correction to the potential in the flat background, and $\textbf{W}_1$ in the curved background (otherwise, in the external gravitational field).
Therefore, below as well as here we separate the parts of the effective potential contributing to the flat case $\textbf{V}$ and to the curved case $\textbf{W}$
\beq
V_{eff}=\textbf{V}+\textbf{W}.
\label{eqVW}
\eeq
At the same time, note that there are no terms in the leading poles/logarithms approximation contributing to $Z(\ph)$ \cite{Iliopoulos:1974ur}.
In addition, it should be noted that the coefficients at the leading poles and logarithms coincide, and this property is typical not only for the first but also for the following loops \cite{Kazakov:2022pkc}. 

\begin{figure}[ht]
 \begin{center}
  \epsfxsize=6cm
 \epsffile{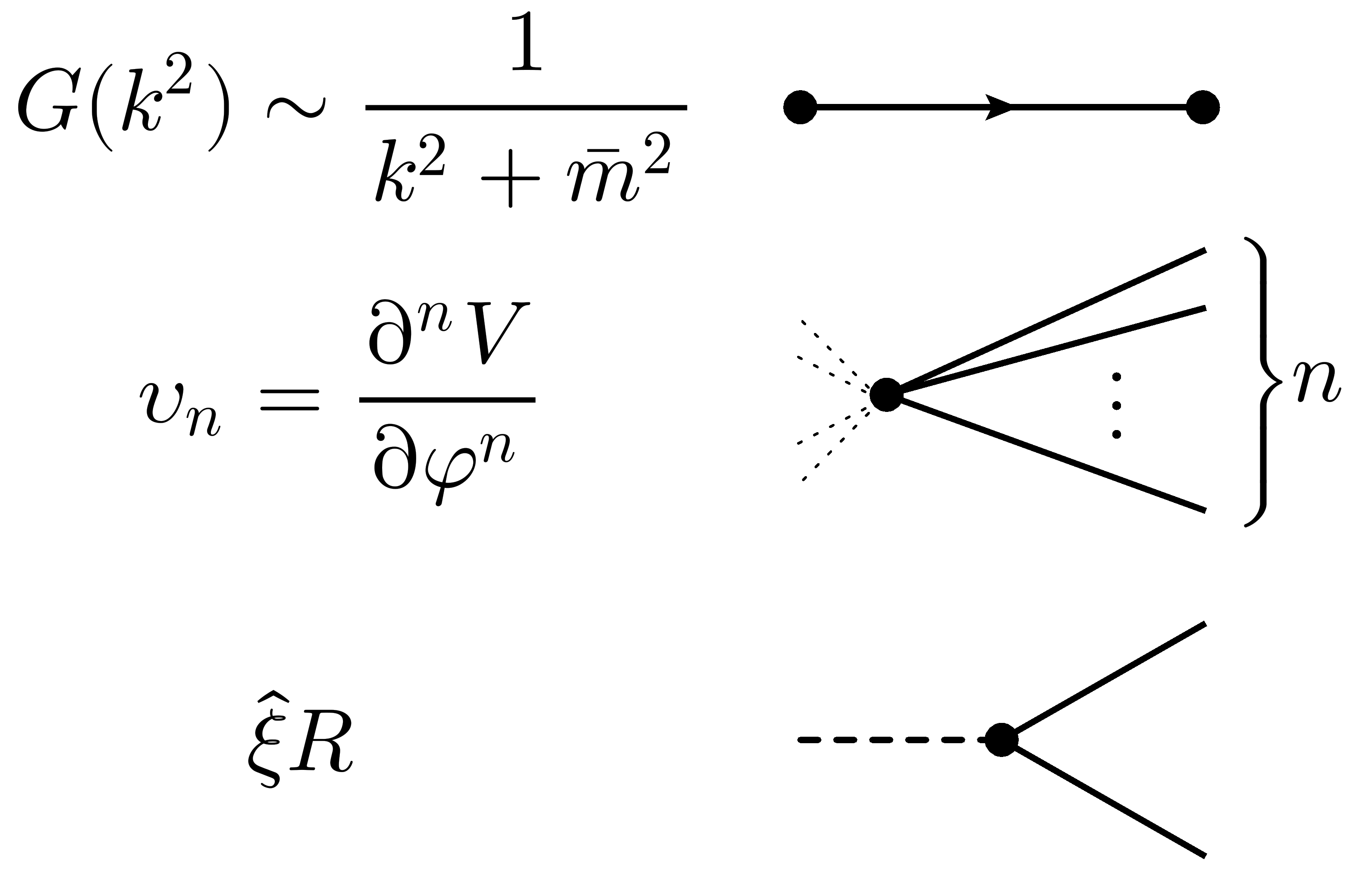}
 \end{center}
 \vspace{-0.2cm}
 \caption{Feynman rules for vacuum $\log$-divergent diagrams on a curved background in the linear approximation by R. The black solid lines denote quantum lines, the dashed line denotes the external field line and the dashed line denotes classical scalar lines} 
\label{fig::effrules}
 \end{figure}

\begin{figure}[ht]
 \begin{center}
  \epsfxsize=7.5cm
 \epsffile{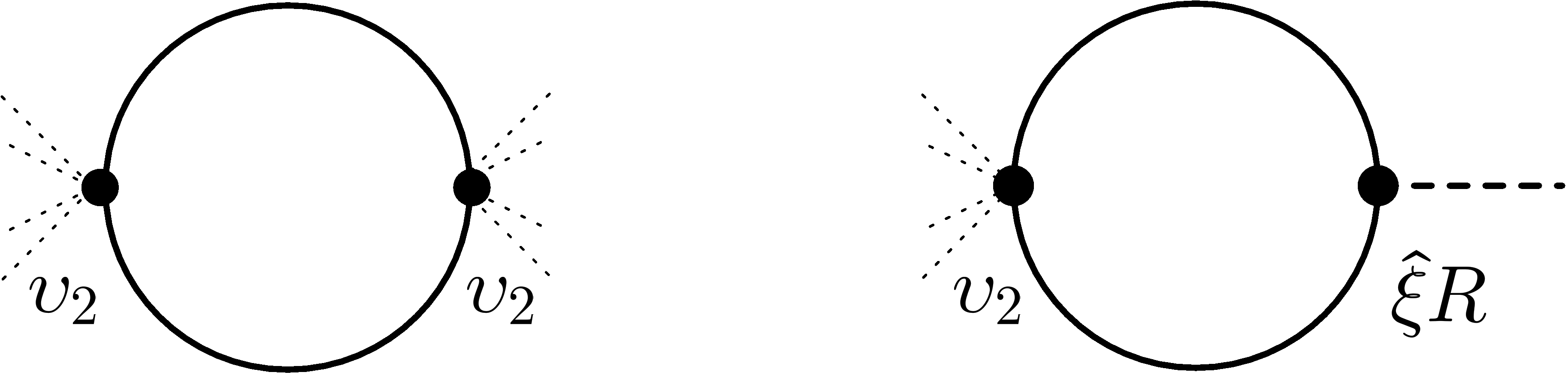}
 \end{center}
 \vspace{-0.2cm}
 \caption{One-loop diagrams corresponding to \eqref{1loopflat} and \eqref{1loopcurved}} 
\label{fig::eff1loop}
 \end{figure}

One can notice that the leading divergences in the corrections coincide with the divergences of the diagrams constructed with Feynman's rules in the external field shown in Fig. \ref{fig::effrules}, and their coefficients correspond to the symmetry factors of each of the diagrams in Fig. \ref{fig::eff1loop} (this is how it is easier to extract  $\log$-divergent parts of the effective potential).

Exploiting this equivalence, one can easily write out and calculate the leading poles for two-loop and three-loop diagrams. For two loops we have two types of singular contributions:
\beq
\Delta V_2= \Delta \textbf{V}_2+ \Delta\textbf{W}_2,
\eeq
where
\beq
\Delta \textbf{V}_2=\frac{1}{8\epsilon^2} v_4 v_2^2+\frac{1}{8\epsilon^2} v_3^2 v_2,
\eeq
and
\beq
\Delta \textbf{W}_2=\frac{1}{8\epsilon^2}v_3^2\hat{\xi}R+\frac{1}{4\epsilon^2}   v_2 v_4 \hat{\xi}R.
\eeq
For three-loop diagrams, the result is also expressed as the sum of 
\beq
\Delta V_3=\Delta\textbf{V}_3+\Delta\textbf{W}_3,
\eeq
where
\beq
\Delta\textbf{V}_3=\frac{1}{48\epsilon^3}v_3^4+\frac{3}{16\epsilon^3} v_2 v_4 v_3^2+\frac{1}{8\epsilon^3} v_2^2 v_5 v_3+\frac{5}{48\epsilon^3} v_2^2 v_4^2+\frac{1}{48\epsilon^3} v_2^3 v_6,
\eeq
and also
\beq
\Delta\textbf{W}_3=\frac{1}{16\epsilon^3} \hat{\xi}  R v_6 v_2^2+\frac{5}{24\epsilon^3} \hat{\xi}  R v_4^2 v_2+\frac{1}{4\epsilon^3} \hat{\xi}  R v_3 v_5 v_2+\frac{3}{16\epsilon^3} \hat{\xi} R v_3^2 v_4.
\eeq
 \begin{figure}[ht]
 \begin{center}
  \epsfxsize=12cm
 \epsffile{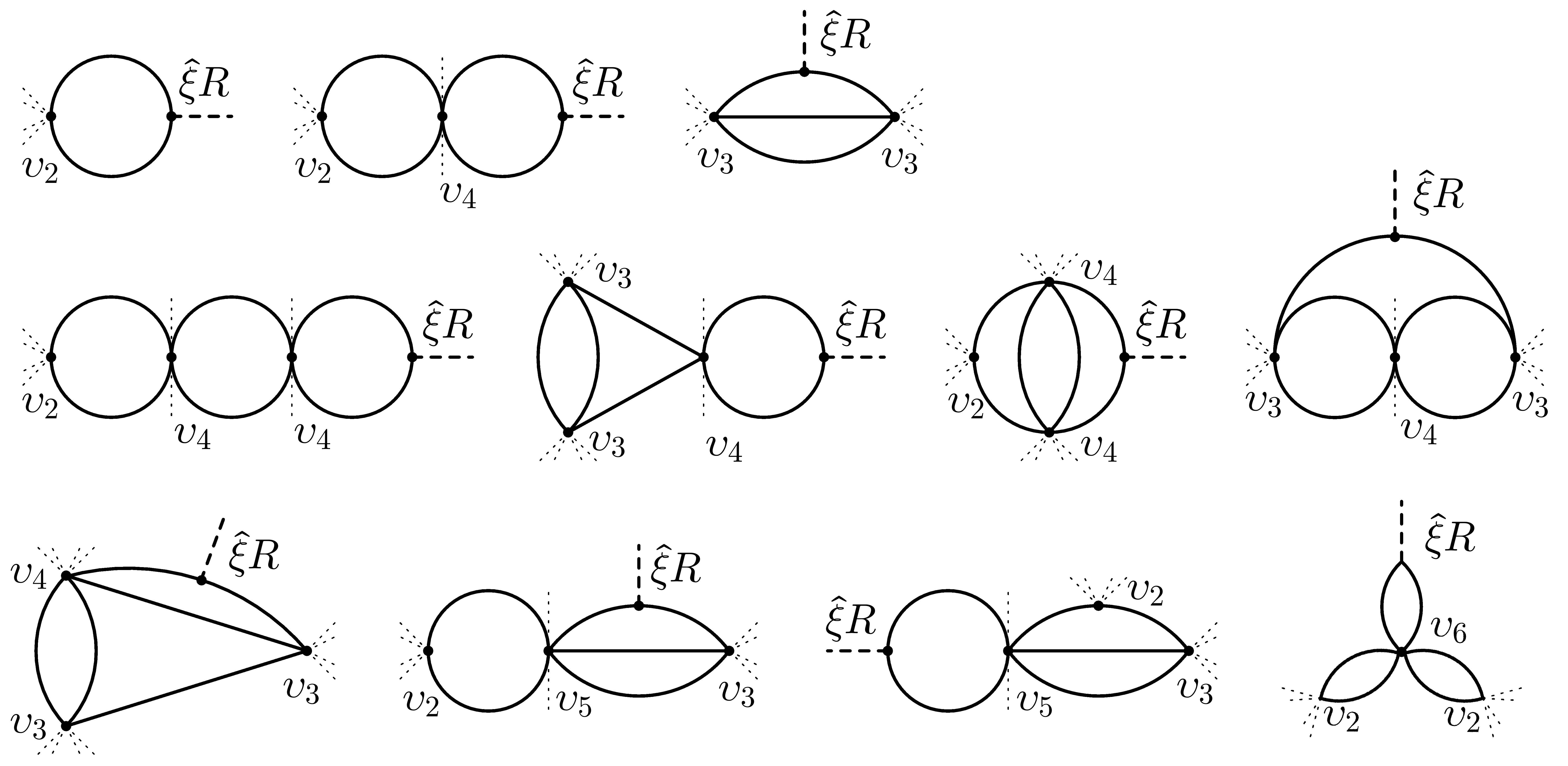}
 \end{center}
 \vspace{-0.2cm}
 \caption{The Feynman diagram series giving contributions to $\Delta\textbf{W}_n$. They are shown here up to three-loop order. } 
\label{fig::eff3loops}
 \end{figure}
The series of the coefficients $\Delta \textbf{V}_n$ is the same series studied in detail in Ref. \cite{Kazakov:2022pkc}. Therefore, below we focus on the recurrence equations for the coefficients  $\Delta \textbf{W}_n$.

\section{Renormalisation group equations}

\begin{figure}[ht]
 \begin{center}
  \epsfxsize=11cm
 \epsffile{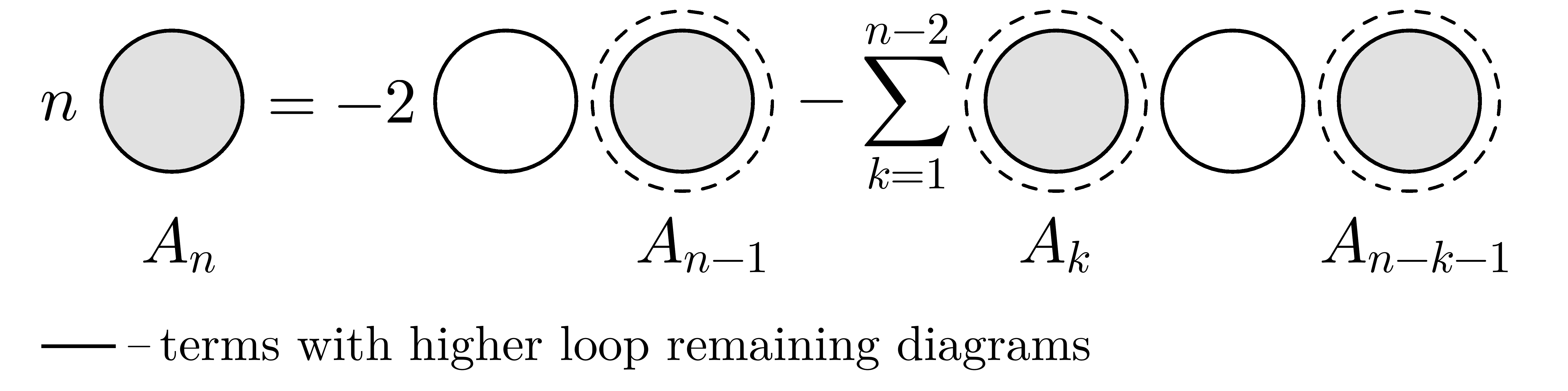}
 \end{center}
 \vspace{-0.2cm}
 \caption{Reccurrence relations for the leading singularities as a result of $\cal{R}'$-operation.} 
\label{fig::Rop}
 \end{figure}

It is known that the ${\mathcal{R}}$ operation~\cite{BogoliubovBook,Collins} applied to a diagram with $n$-loops subtracts first the ultraviolet divergences in the subgraphs starting at one loop and ending at $(n-1)$-loops, and then subtracts the remaining $n$-loop divergence, which is bound to be local by the Bogoliubov-Parasiuk theorem \cite{BP,Hepp,Zimmermann}. This $n$-loop pole left after the incomplete ${\mathcal{R}}$-operation ($\mathcal{R}'$-operation) is exactly the object we need.
The locality requirement states that the leading $1/\epsilon^n$ divergence in $n$-loops of $A^{(n)}_n$ is given as follows
~\cite{Kazakov:2018zgi}:
\beq
n A^{(n)}_n=(-1)^{n+1} A^{(1)}_n, \label{leadA}
\eeq
where $A^{(1)}_n$ is the one-loop divergence remaining after subtracting the $(n-1)$-loop counterterm as a result of the incomplete ${\cal R'}$-operation \cite{Vasiliev}. Note also that the result of the ${\cal R'}$-operation applied to the $n$-loop graph is the recurrence formula \cite{Kazakov:2022pkc} depicted in Fig \ref{fig::Rop}.
Using this expression and applying it in the same way as in \cite{Kazakov:2022pkc}, it is possible from the expression for the coefficient at the pole of the first loop to obtain the coefficient for the second and so on. For example, for the first diagram
\beq
2\frac{1}{\epsilon^2}v_4 v_2\hat{\xi}R ~A^{(2)}_2=-2\times\frac{1}{\epsilon}~v_4 v_2\hat{\xi}R~\frac{\left(A_{1}^{(1)}\right)^2}{\epsilon},
\eeq
and for the second one
\beq
 2\frac{1}{\epsilon^2}v_3^2 \hat{\xi}R ~A^{(2)}_2=-\frac{1}{\epsilon}~v_3^2\hat{\xi}R~\frac{\left(A_{1}^{(1)}\right)^2}{\epsilon},
\eeq
it is known that the first loop coefficient $A_1^{(1)}=1$ can be derived taking into account the symmetry factor, so $$\Delta\textbf{W}_{2}=\frac{1}{8\epsilon^2}v_3^2\hat{\xi}R+\frac{1}{4\epsilon^2}   v_2 v_4 \hat{\xi}R.$$ 
With the help of the specified Feynman rules in Fig.\ref{fig::effrules}, one can continue to write out related expressions for the diagrams in Fig. \ref{fig::eff3loops}, giving the contribution to the non-minimal interaction term.

The obtained relations allow us to write the recurrence equation for the connection of coefficients at higher poles for the first series of diagrams. It can be written for the 'flat' diagram series as \cite{Kazakov:2022pkc}
\beq
n \Delta \textbf{V}_n=\frac{1}{4} \sum_{k=0}^{n-1} \dfrac{\pa^2}{\pa \ph^2} \Delta \textbf{V}_{k} \dfrac{\pa^2}{\pa \ph^2}\Delta \textbf{V}_{ n- k-1}\label{VRGrec},
\eeq
and for the 'curved' ones 
\beq
n \Delta \textbf{W}_n=\frac{1}{2} \sum_{k=0}^{n-1} \dfrac{\pa^2}{\pa \ph^2} \Delta  \textbf{W}_{k} \dfrac{\pa^2}{\pa \ph^2}\Delta \textbf{V}_{ n- k-1}\label{WRGrec}.
\eeq
Moreover, if we introduce the variable $z=\la/\epsilon$, the functions summing up all leading singular contributions turn to 
\begin{equation}
\Sigma_\la (z,\ph) = \sum_{n=0}^\infty (-z)^n \Delta \textbf{V}_n, ~~ \Sigma_\xi (z,\varphi) = \sum_{n=0}^\infty (-z)^n \Delta \textbf{W}_n.
\end{equation}
Therefore, equations \eqref{VRGrec} and \eqref{WRGrec} become a system of equations
\begin{equation}
\dfrac{\partial\Sigma_\la}{\partial z} = -\dfrac{1}{4}\left(\dfrac{\partial^2\Sigma_\la}{\partial\varphi^2}\right)^2,
\label{VRG}
\end{equation}
\begin{equation}
\dfrac{\partial\Sigma_\xi}{\partial z} = -\dfrac{1}{2}\dfrac{\partial^2\Sigma_\xi}{\partial\varphi^2}\dfrac{\partial^2\Sigma_\la}{\partial\varphi^2},
\label{WRG}
\end{equation}
with the initial conditions 
$\Sigma_\xi(0,\varphi)=1/2 \,\hat{\xi}R \ph^2$ and $\Sigma_\la(0,\varphi)=V_0(\varphi)$. The first equation is completely consistent with the equation found by \cite{Kazakov:2022pkc}. The found equations also correspond to the effective potential in the leading pole approximation\footnote{Note here again that the constant part of the effective potential and various generalisations of the concept of running cosmological constant by analogy with \cite{Shapiro:2009dh,Sobreira:2011ep} are not discussed within the framework of this work.} since, as it was noted earlier, the leading logarithms and the leading poles have the same coefficients so that the effective potential is given by
\beq
V_{eff}=\lambda \Sigma_\la(z,\ph)- \left(\frac{1}{12} R\ph^2 -~\Sigma_\xi(z,\ph)\right)\bigg|_{z \rightarrow -\frac{\lambda}{16\pi^2}\log(\lambda v_2/\mu^2)}
\label{varchange}.
\eeq
In the next section, we discuss a particular form of the effective potential for a scalar theory with an arbitrary power-like potential.

\section{Power-like potentials}

The simplest example for studying the behavior of equations are the power-like potentials of  $$V_0=\dfrac{\ph^p}{p!}.$$ It is easy to check the validity of the obtained solutions of equations with the help of these potentials. For convenience, we can introduce a dimensionless variable $y = z \ph^{p-4}$ that can be tracked in the loop expansion (due to obvious dimensional reasons, the renormalisable theory is only where $y=z$) and substitute the following ansatz:
\begin{equation}
\begin{split}
&\Sigma_\la(z,\varphi) = \dfrac{\varphi^p}{p!}f_1(y),\\
&\Sigma_\xi(z,\varphi) = \hat{\xi} R \dfrac{\varphi^2}{2!}f_2(y).
\label{sigma_split}
\end{split}
\end{equation}
Then the system of equations (\ref{VRG}-\ref{WRG}) has the following form of non-linear ordinary differential equations (ODEs):
\begin{equation}
-4p!f_1'(y) = \left[ p (p-1) f_1(y) + (p-4)(3p-5) y f_1'(y) + (p-4)^2 y^2 f_1''(y) \right]^2,
\label{vrgfp}
\end{equation}
\beq
\begin{split}
-2 p! f_2'(y)=\left[(p-4) y \left((3 p-5) f_1'(y)+(p-4) y f_1''(y)\right)+(p-1) p f_1(y)\right] \times \\ \times \left[(p-4) y \left((p-1) f_2'(y)+(p-4) y f_2''(y)\right)+2 f_2(y)\right]\label{wrgfp},
\end{split}
\eeq
with the initial conditions, respectively, 
\beq
\begin{split}
    &f_1(0)=1,~ f_1'(0)=-\frac{p(p-1)}{4 (p-2)!},\\
&f_2(0)=1,~  f_2'(0)=-\frac{1}{(p-2)!}.
\end{split}
\eeq
The form of the system suggests that the simplest form of the equation is obtained when $p=4$, which recovers the limit of renormalisable massless model. 

It is worth noting the applicability conditions imposed on the parameters of the theory. So one of them is the condition of smallness of the expansion parameter $|y|\ll 1$, but it is also necessary to satisfy the condition of the leading logarithmic approximation. Explicitly, these conditions take the following form:
\beq
\dfrac{\lambda}{16 \pi^2}\varphi^{p-4} \ll 1
\label{ineq_per},
\eeq

\beq
\ln{ \dfrac{\lambda \varphi^{p-2}}{(p-2)! \mu^2} }\gg 1
\label{ineq_LL}.
\eeq
The conditions \eqref{ineq_per}, \eqref{ineq_LL} can be satisfied by considering small values of the interaction constant $\lambda$ and the transmutation parameter $\mu$. Indeed, substituting inequality \eqref{ineq_per} into \eqref{ineq_LL}
\beq
1 \ll \ln{ \dfrac{\lambda \varphi^{p-2}}{(p-2)! \mu^2} } \ll \ln{ \dfrac{ (16 \pi^2)^{ \frac{p-2}{p-4} } }{ (p-2)! \lambda^{\frac{2}{p-4}} \mu^2} },
\eeq
we can get the desired restrictiion on $\lambda$ and $\mu$
\beq
\lambda^{\frac{2}{p-4}} \mu^2 \ll \dfrac{ (16 \pi^2)^{ \frac{p-2}{p-4} } }{ (p-2)! }.
\eeq
That is, only if these conditions are satisfied, the functions $\Sigma_\la$ and $\Sigma_\xi$ for power-like potentials are convergent and the approximation of the leading logarithms for the effective potential is valid. In the following subsections, we analyse in detail all possible behaviors for the theory with self-interaction $\ph^p$.

\subsection{Renormalisable $p=4$ case}
With this choice of the $p=4$ potential, the system of equations can be expressed as 
\begin{equation}
\begin{split}
&\Sigma_\la(z,\varphi) = \dfrac{\varphi^4}{4!}f_1(z),\\
&\Sigma_\xi(z,\varphi) = \hat{\xi} R \dfrac{\varphi^2}{2!}f_2(z),
\end{split}
\end{equation}
so the equations are re-expressed as
\begin{equation}
f_1'(z) =-\dfrac{3}{2}f_1^2(z),
\label{vrgf4}
\end{equation}
\beq
\begin{split}
f_2'(z)=-\dfrac{1}{2}f_1(z) f_2(z),
\end{split}
\eeq
and solutions can be obtained
\beq
\begin{gathered}
    f_1(z)=\frac{1}{1+\frac{3}{2}z},~ f_2(z)=\left(1+\frac{3}{2}z\right)^{-1/3},
\end{gathered}
\eeq
and in the number $3/2$ one can recognize the one-loop beta function in the renormalisable scalar theory $\ph^4$.
After substituting \eqref{varchange}, we arrive at the resummed effective potential of the form 
\beq
V_{eff}=\frac{\la\ph^4}{4!}\frac{1}{1-\frac{3}{2}\frac{\la}{(4\pi)^2}\log\left(\frac{\la \ph^2}{2\mu^2}\right)}-\frac{1}{2}R\ph^2 \left[\frac{1}{6}-\hat{\xi}\left(1-\frac{3}{2}\frac{\la}{(4\pi)^2}\log\left(\frac{\la \ph^2}{2\mu^2}\right)\right)^{-1/3}\right].
\eeq
This result is fully consistent with the RG-summed results in the renormalisable model known in the literature \cite{Ishikawa:1983kz,Toms:1982af,Inagaki:2014wva}. Below we compare the behavior of this expression with the effective potential for non-renormalisable models

\subsection{Qualitative analysis for $p>4$}
In this section, we expand the resulting  RG equations near the critical dimension $p=4$.  For this purpose it is necessary to rewrite the equation for the theory with a power-like potential with $p=4+\delta$:
\beq
\begin{gathered}
-4 (4+\delta)! ~f'_1(y)=\left[y \delta  \left((3 \delta +7) f_1'(y)+y \delta  f_1''(y)\right)+(\delta +3) (\delta +4) f_1(y)\right]^2 
\end{gathered}
\eeq

\beq
\begin{gathered}
-2 (4+\delta)! ~f'_2(y)=\left[y \delta  \left((3 \delta +7) f_1'(y)+y \delta  f_1''(y)\right)+(\delta +3) (\delta +4) f_1(y)\right] \\ \left[y \delta  \left((\delta +3) f_2'(y)+y \delta  f_2''(y)\right)+2 f_2(y)\right],
\end{gathered}
\eeq
and at $\delta\ll1$ in the linear approximation by $\delta$ the ODE system takes the following form:
\beq
4\Gamma(5+\delta)f_1'(y)=-6 f(y) \left(7 y \delta  f_1'(y)+(7 \delta +6) f_1(y)\right)
\label{f_1_lin}
\eeq
\beq
2\Gamma(5+\delta)f_2'(y)=-7  y \delta  f_2(y) f_1'(y)- f_1(y) \left(18 y \delta  f_2'(y)+(7 \delta +12) f_2(y)\right)
\label{f_2_lin}.
\eeq
It is clear from the form of the equation itself that, for example, the notion of a beta function for non-renormalisable theories is irrelevant, since the renormalisation of even leading divergences can be carried out in them only by means of sophisticated operators \cite{Kazakov:2018zgi}.
It is still difficult to solve this system of equations in analytically closed form, but this representation allows us to perform a numerical analysis of the qualitative behavior of these equations and the general properties of their solutions in renormalisable and non-renormalisable regimes.

 \begin{figure}[ht]
 \begin{center}
  \epsfxsize=7.5cm
 \epsffile{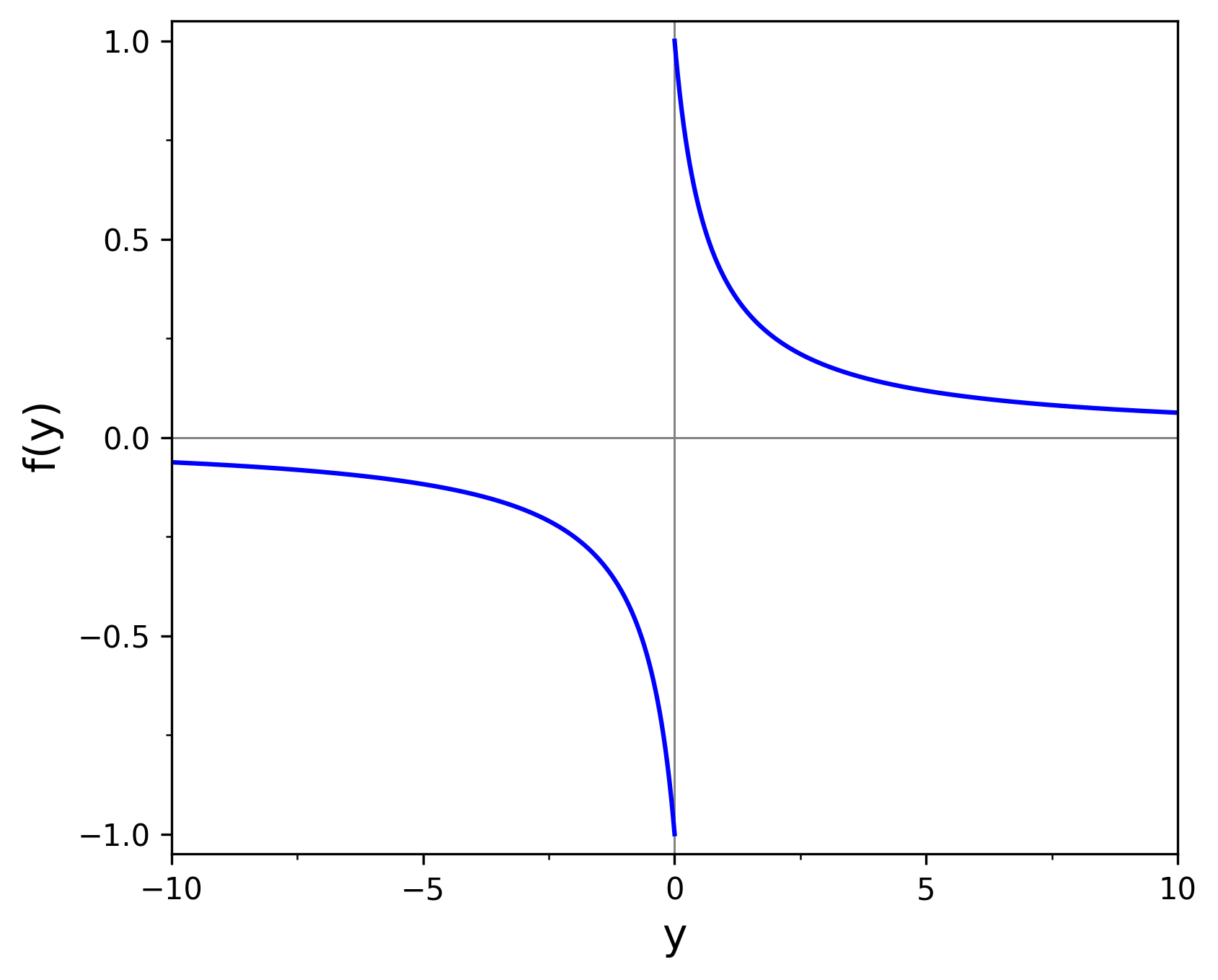}
 \end{center}
 \vspace{-0.2cm}
 \caption{General form of the solutions $f_1$ and $f_2$ for equations \eqref{vrgfp} and \eqref{wrgfp}; $y=0$ corresponds to discontinuity.} 
\label{fig::generalF}
 \end{figure}

It should be noted that \eqref{f_1_lin} in the linear approximation is equivalent to the solution of
\beq
f_1(y)^{\frac{6}{7 \delta +6}}+q y f_1(y)=1 \label{quadsol},
\eeq
where $q=24/\Gamma (\delta +5)$. It is easy to see that \eqref{quadsol} recovers the solution as a geometric progression in the limit $\delta \rightarrow 0$. This equation is sufficient to determine the features of the solutions for non-renormalisable models. For illustration, we give the simplest nontrivial solution of equation \eqref{quadsol} which is achieved when $\delta=3/7$, so it turns to
\beq
f_1^3 q^3 y^3+f_1^2 \left(1-3 q^2 y^2\right)+3 f_1 q y-1=0. \label{eqcube}
\eeq
This equation, like any cubic equation, is easily solvable with respect to $f_1$ but is explicitly representable only in a cumbersome form, which we do not give here. Nevertheless, we must emphasise that the solution of this equation contains a discontinuity at $y=0$, since the right and left limits of the function do not coincide $\lim_{y\to\pm0}f_1(y)=\pm1$ (this feature can be seen directly from equation \eqref{eqcube}). We can conclude that all the following solutions for non-renormalisable models have a similar behaviour near the zero of the argument. This observation confirms the solutions with discontinuities studied in \cite{Kazakov:2022pkc,Iakhibbaev:2024fjf}. The general properties of the behaviour of equations of the type (\ref{vrgfp}-\ref{wrgfp}) obviously need a more detailed and general group-theoretic analysis.

The second equation \eqref{f_2_lin} integrates easily if the solution of \eqref{f_1_lin} is known, and it has the following form:
\beq
f_2(y)=\exp{\int_{0}^{y} d\tau ~\frac{2 \left((12 +7  \delta)  f_1(\tau )+7 \delta \tau  f_1'(\tau )\right)}{l-36 \tau  \delta  f_1(\tau )}},
\eeq
where $l={\Gamma(5+\delta)}$.
It is clear that even for not the most complicated solution for $f_1$ one cannot obtain an explicit analytic expression for $f_2$ from \eqref{quadsol}. Therefore, in the following we only give the numerical results for $f_1$ and $f_2$ in the limit of the $\delta$ linear approximation and the corresponding $V_{eff}(\ph)$ as well as numerical solutions of some equations \eqref{vrgfp} and \eqref{wrgfp}.
 
\begin{figure}[ht]
    \begin{center}
        \begin{minipage}{0.5\textwidth}
            \centering
            \epsfxsize=8.0cm
            \epsffile{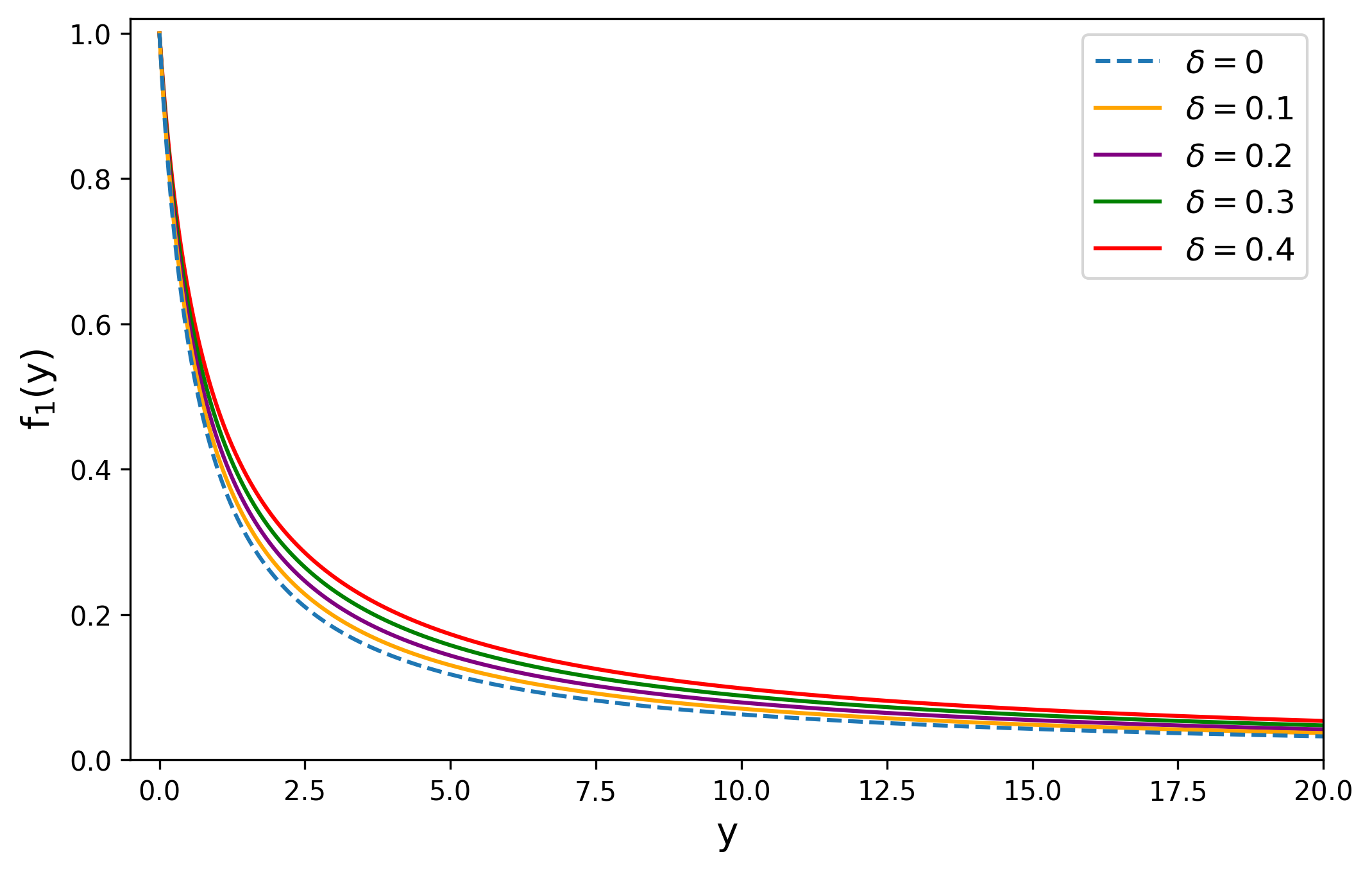}
            \label{fig::f1lin}
        \end{minipage}
        \hfill
        \begin{minipage}{0.5\textwidth}
            \centering
            \epsfxsize=8.0cm
            \epsffile{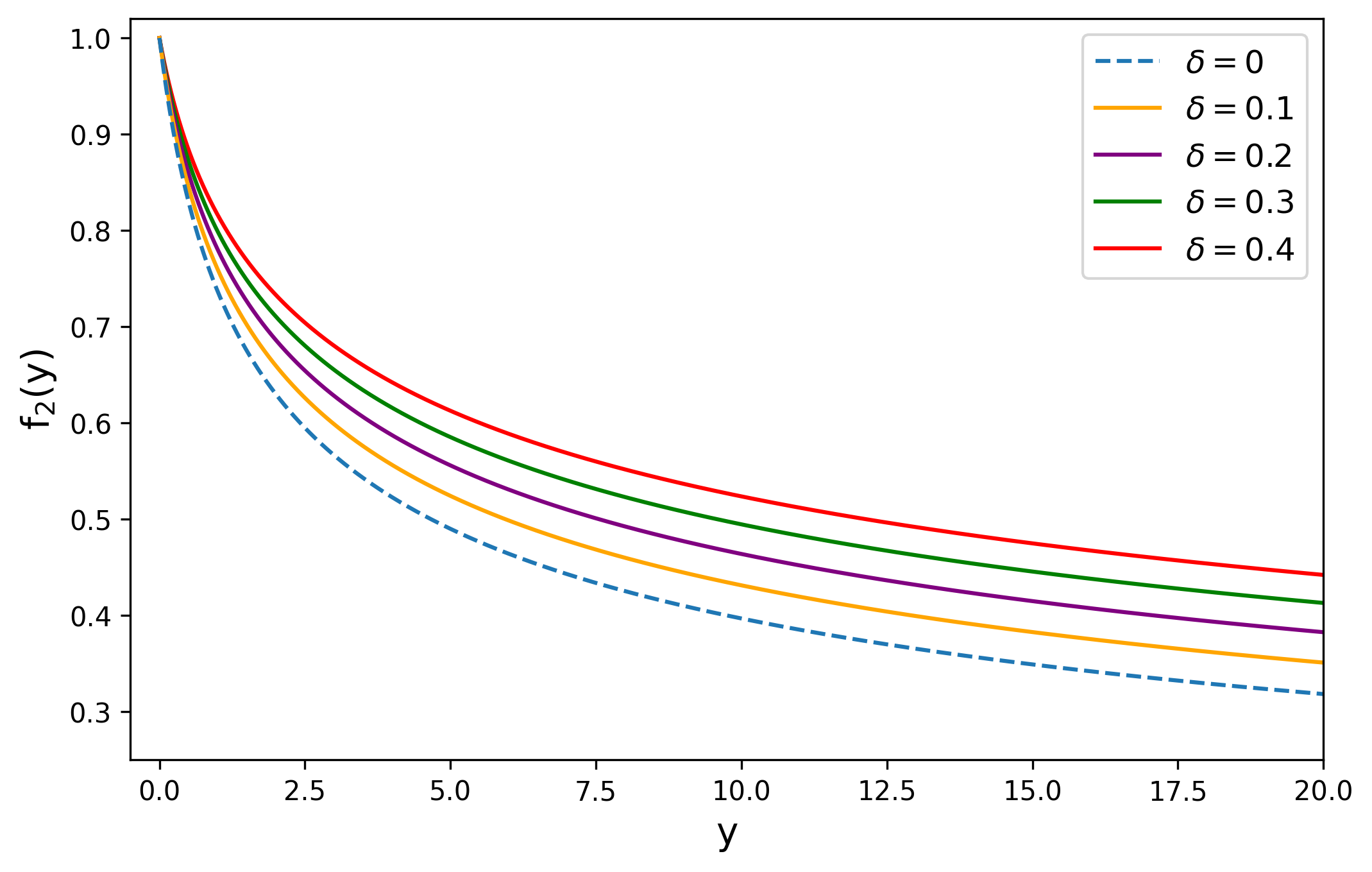}
            \label{fig::f2lin}
        \end{minipage}
    \end{center}
    \vspace{-1.cm}
    \caption{Solutions $f_1(y)$ and $f_2(y)$ for different $\delta$}
    \label{fig::f_1_2_lin_figs}
    \vspace{-0.2cm}
\end{figure}

Figure \ref{fig::f_1_2_lin_figs} shows the solutions of the functions $f_1(y)$ and $f_2(y)$ for the system of ODEs \eqref{f_1_lin},\eqref{f_2_lin} in the linear approximation by $\delta$. It is worth noting that these functions decrease rapidly as $y\rightarrow\infty$, i.e., as $\epsilon \rightarrow0$ the functions tend either to a small constant or to zero. This means that when the regularisation is removed, the functions $f_1$ and $f_2$ turn to small constants or zero. As $\delta$ increases, the solutions decrease more slowly.

With these functions one can obtain a direct effective potential. It is obvious that in the case of conformal coupling $\xi=1/6$ the effects of quantum corrections will not be observed in $V_{eff}(\ph)$. The situation changes in the minimal coupling $\xi=0$. 

\begin{figure}[ht]
    \begin{center}
    \begin{minipage}{0.3\textwidth}
            \epsfxsize=5.0cm
            \epsffile{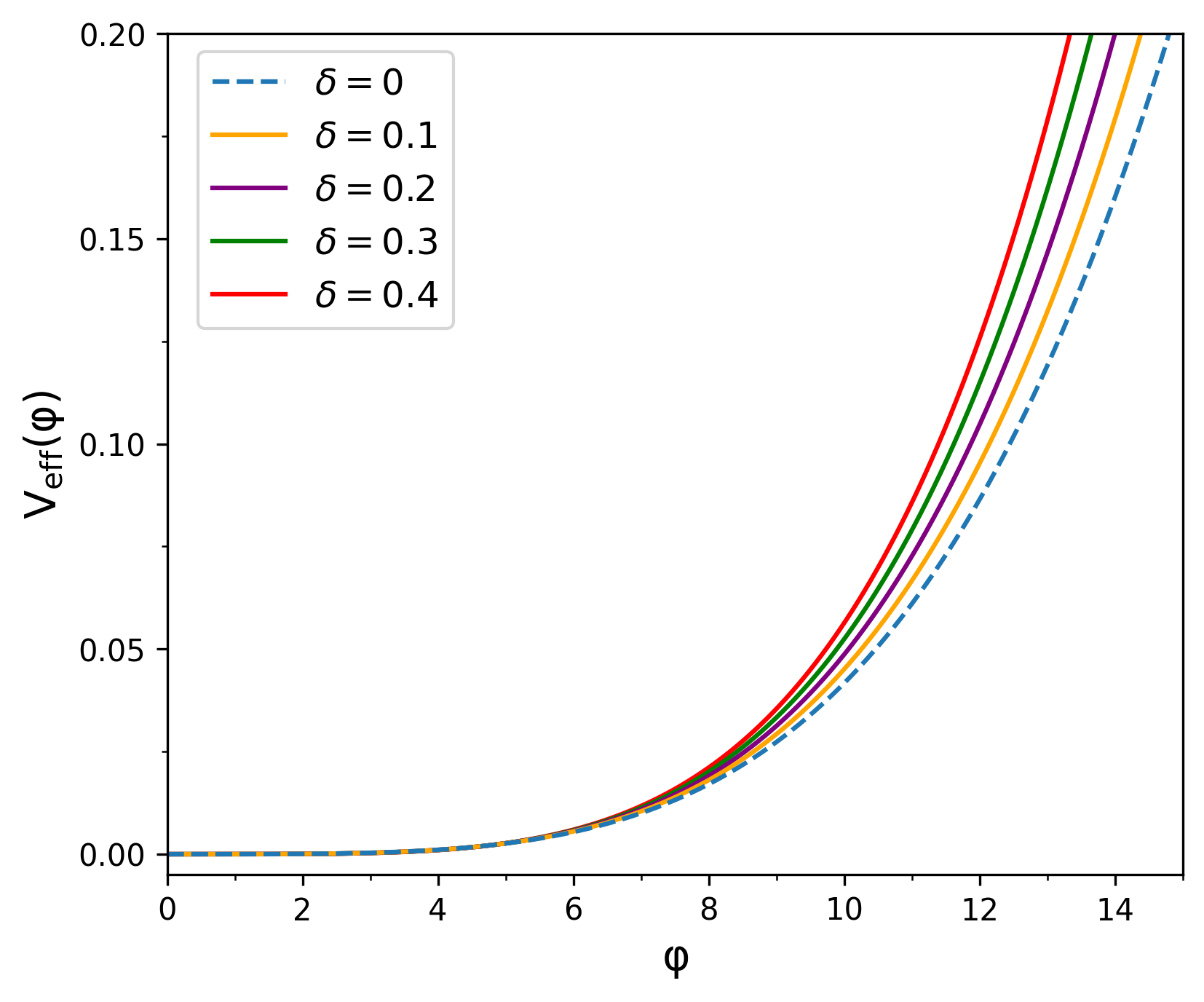}
            \label{fig::V_R}
        \end{minipage}
        \hfill
        \begin{minipage}{0.3\textwidth}
            \epsfxsize=5.0cm
            \epsffile{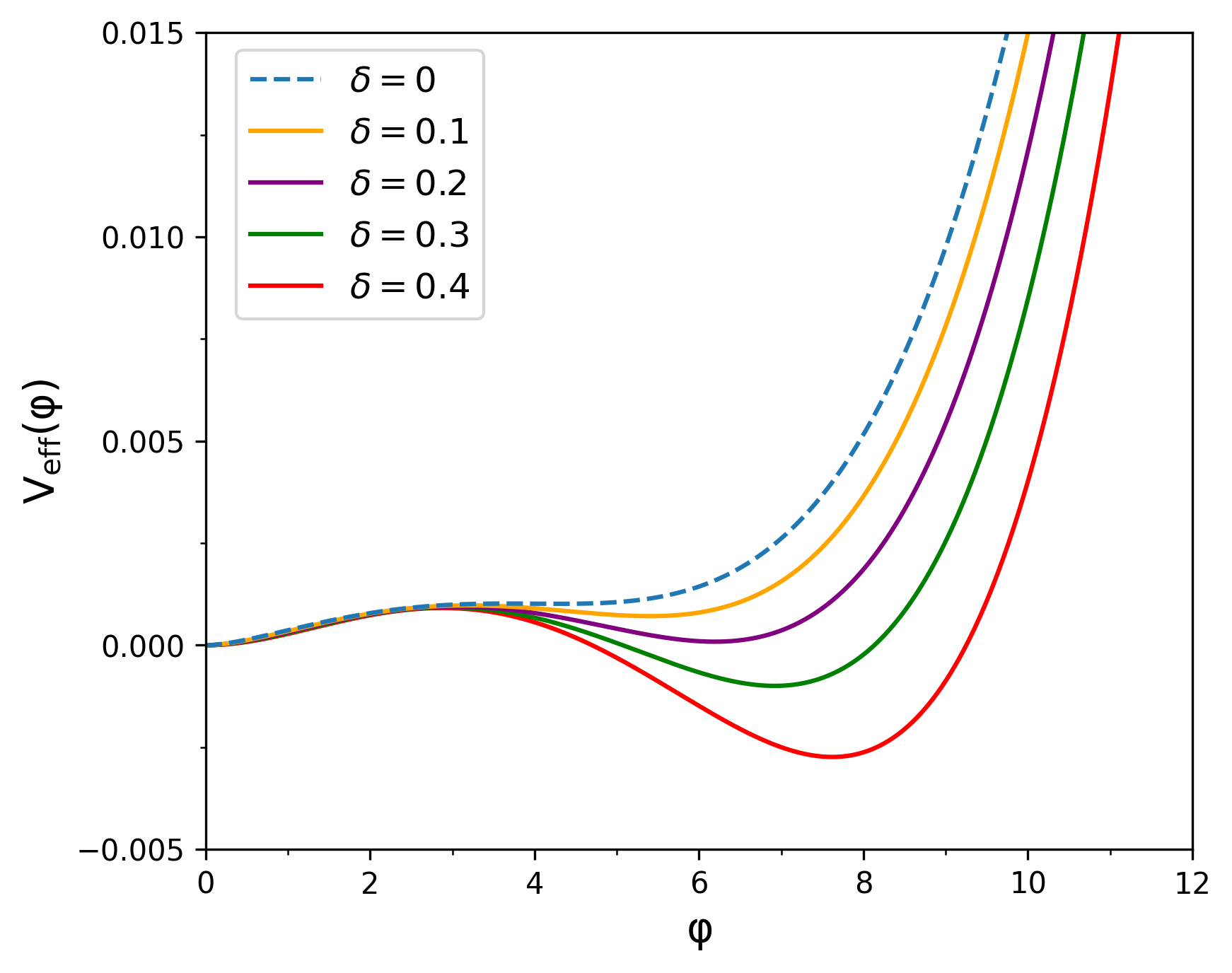}
            \label{fig::V_RC1}
        \end{minipage}
        \hfill
        \begin{minipage}{0.3\textwidth}
            \epsfxsize=5.0cm
            \epsffile{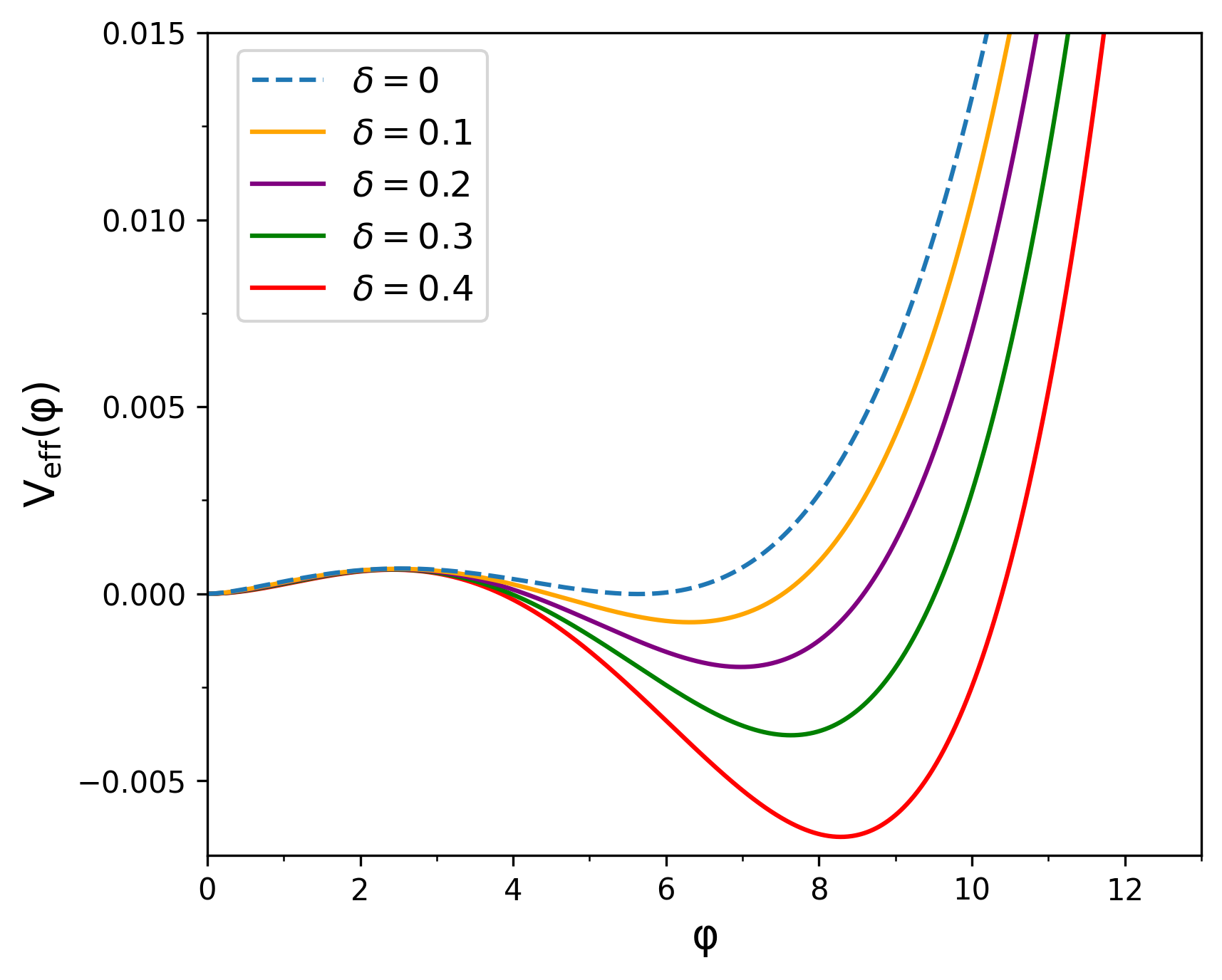}
            \label{fig::V_RC2}
        \end{minipage}
    \end{center}
    \vspace{-0.9cm}
    \caption{Effective potential $V_{eff}(\ph)$ for $\xi=0$ and different critical values of the curvature $R$ as the curvature increases from $0$ up to $R_{C2}$ with $\delta \in [0,0.4]$. The parameters chosen for illustrative purposes are $R_{C1}\sim -10^2 \mu^2$, $R_{C2} \sim -10^4\mu^2$, $\lambda \sim 10^{-4}$.}
    \label{fig::V_eff_RC1_RC2}
\end{figure}

Figure \ref{fig::V_eff_RC1_RC2} depicts the effective potential $V_{eff}$ in the linear approximation by $\delta$ for different conditional values of the curvature $R$. The distinguished cases for the effective potential with $\delta=0$ and critical values of the curvature $R=0, R=R_{C1}, R=R_{C2}$ are depicted as some basis where the critical values of the curvature $R_{C1}$ and $R_{C2}$ are responsible for the appearance of the first additional local minimum and the transformation of this minimum from local to global, respectively.
Here the values are conventional, since, for example, it is clear that for large curvatures $|R_{C2}| \gg v_2$ the weak gravity approximation does not hold. These examples show that at large fields $\ph$ there is the symmetry breaking and  formation of new minima for a sufficiently large $R$ arising due to the fact that the contribution of the non-minimal interaction term becomes comparable to the effective potential for $V(\ph)$. It is also obvious here that the effect of the ''curved part'' of the effective potential in non-renormalisable theories has a stronger infuence than in the renormalisable ones. Thus, the level of the additional minimum due to quantum corrections lowers as the $\delta$ parameter increases, but the qualitative behavior of the effective potential is preserved.

\section{Cosmological implications}

Let us now to turn to the consequences of the potentials considered above for the inflationary stage and impact of quantum effects of gravity-scalar interaction on the evolution of the early Universe. We rewrite the action of \eqref{action} taking into account \eqref{eqVW} in the following form:
\beq
S[g_{\mu \nu},\ph]=\int d^4x \sqrt{-g} \left(\dfrac{1}{2} F(\ph) R + \dfrac{1}{2}g_{\mu \nu}\partial^\mu\varphi\partial^\nu\varphi - \textbf{V}(\ph)\right),
\eeq
where the function $F(\ph)$ is given
\beq
F(\ph) = 1  - \dfrac{2}{R} \textbf{W}(\ph).
\eeq
It is implied here that $\ph$ is the scalar field of inflaton, $\textbf{V}(\ph)$ is the ''flat'' part of the scalar potential, and $\textbf{W}(\ph)$ is the ''curved'' part. Note that the dependence on the Ricci scalar is reduced in the final result for the RG-resummed functions.

A good approximation to describe the accelerated expansion of the Universe is the slow-roll approximation. The slow-roll conditions are charaterised by the Hubble flow parameters $\bar{\epsilon}, \bar{\eta}$ etc. (see for details \cite{Liddle:1994dx, Martin:2013tda, weinberg2008cosmology}). Due to the presence of non-minimal scalar-tensor interaction, one can extract them using the additional function
\beq
\mathcal{K}\left(\ph\right) = \frac{1}{F}+\frac{3}{2}\left(\frac{F_{,\ph}}{F}\right)^{2},
\eeq
as it was done in \cite{Karciauskas:2022jzd}
\footnote{There are more precise ways to determine the Hubble parameters in models with non-minimal interactions (see ref. \cite{Pozdeeva:2025ied}), but in the present paper we work with the simplest ones.}. 
Folowing the same work, one can obtain the first slow-roll parameter as follows:
\beq
\bar{\epsilon} = \frac{1}{2\mathcal{K}}\left(\frac{\textbf{V}_{,\ph}}{\textbf{V}}-2\frac{F_{,\ph}}{F}\right)^{2},
\eeq
meanwhile the second one can be expressed as
\beq
\bar{\eta} = \frac{1}{\mathcal{K}}\left[2\frac{F_{,\ph\ph}}{F}-\frac{\textbf{V}_{,\ph\ph}}{\textbf{V}}-2\frac{F_{,\ph}^{2}}{F^{2}}+\frac{\textbf{V}_{,\ph}^{2}}{\textbf{V}^{2}}+\frac{\mathcal{K}_{,\ph}}{2\mathcal{K}}\left(\frac{\textbf{V}_{,\ph}}{\textbf{V}}-2\frac{F_{,\ph}}{F}\right)\right].
\eeq
It is believed that the accelerated expansion corresponds to the value $\bar{\epsilon}<1$ (slow-roll approximation well defined for small values of $\bar{\epsilon}$ and $\bar{\eta}$). Note that inflation ends at $\bar{\epsilon}=1$.

As is known, the inflationary theory gained its main recognition after it succeeded in explaining the inhomogeneity and anisotropy of cosmic microwave background radiation (CMB). The tensor/scalar perturbations ratio of the expanding Universe and the tensor spectral index extracted from the CMB spectrum are not independent parameters, so that both are determined by the equation of state during inflation which can be expressed through the aforementioned Hubble flow parameters.  Thus, the CMB tilt of scalar perturbations can be obtained as \cite{Martin:2013tda,Karciauskas:2022jzd,Inagaki:2014wva}
\beq
n_{\mathrm{s}}-1 \simeq -2\left(\bar{\epsilon} +\bar{\eta}\right),
\eeq
and also the tensor-to-scalar ratio has the following form:
\beq
r \simeq 16\bar{\epsilon}.
\eeq
The inflationary phase should last approximately $N_e\simeq50-60$ e-foldings, so first we estimate
\beq
N_e  \simeq  \intop_{\ph}^{\ph_{\mathrm{end}}}\frac{\mathcal{K}}{2\frac{F_{,\ph}}{F}-\frac{\textbf{V}_{,\ph}}{\textbf{V}}}\mathrm{d}\ph.
\eeq
and then re-express numerically this parameter with the help of $\bar{\epsilon}$ and $\bar{\eta}$ \cite{Martin:2013tda}. Using these parameters, we can establish the consistency of the above effective potentials in the leading limit with the cosmological observational data.

\begin{figure}[ht]
 \begin{center}
  \epsfxsize= 12 cm
 \epsffile{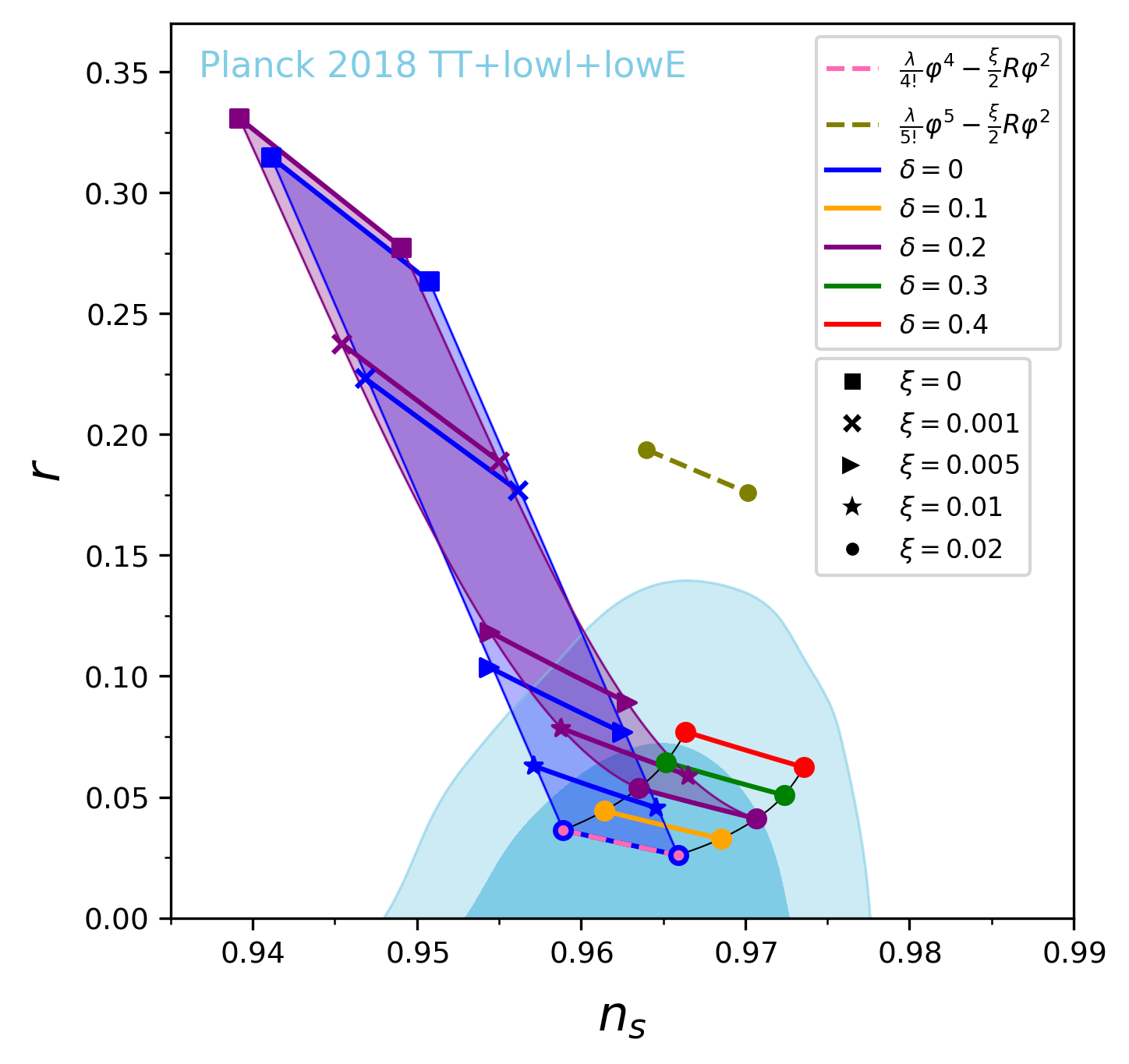}
 \end{center}
 \vspace{-0.5cm}
 \caption{Example of the cosmological observables $(n_s,r)$ for different resummed and not resummed scalar potentials for various $\xi$-coupling values in the $N_e=50$ (left point) and $60$ (right point) e-foldings and at the same fixed values of $R$ and $\mu^2$. } 
\label{fig::nsrRpx}
 \end{figure}

Numerically, we can estimate the duration of inflation for the resulting potentials and determine the behaviour of the cosmological observables $n_s$ and $r$ for the RG-resummed potentials using the Planck/BICEP results \cite{Planck:2018jri,BICEP:2021xfz} . 
Figure \ref{fig::nsrRpx} shows the calculated parameters in the $({n_s,r})$-diagram  for the initial models $\la \ph^4$ and $\la \ph^5$ as well as for the resummed ones with different parameters $\delta$ and $\xi =0.02$. It can be seen that the resummed case for $\delta=0$ almost coincides with the initial model $\la \ph^4$. However, for other cases with non-zero values of $\delta$ (which corresponds to non-renormalised models) the cosmological parameters $n_s$ and $r$ slowly increase, and there is a gradual exit from the confidence zone. Nevertheless, it can be seen that it is possible to fit the model in the area of confident values so that they are still within it for the fixed plot parameters even for models with higher $\delta$. 

Also for demonstration purposes, the blue and purple regions corresponding to $\delta=0$ and $\delta=0.2$, respectively, for different values of $\xi$ are plotted. It should be noted that as the parameter $\xi$ increases, the initial case with $\delta=0$ is deeper in the confidence region of the observed data (i.e., with smaller $r$). Meanwhile, the subsequent shift for non-renormalisable models with nonzero $\delta$ is larger, which may lead to a larger shift away from the confidence zone. 
In turn, decreasing the parameter $\xi$ leads to the initial case with $\delta=0$ with larger values of $r$, but to a smaller spread for cases with $\delta \neq 0$.

\section{Discussion}
We have shown that quantum corrections to the effective potential can be calculated for an arbitrary initial classical potential without renormalisability constraint even for a scalar-tensor theory with non-minimal conformal interaction in the linear curvature limit.  Certainly, this can be done by assuming that the ultraviolet divergences are removed by some subtraction procedure. We do not discuss here the problem of arbitrariness of subtraction methods but derive corollaries from the independence of the leading logarithmic approximation from this unknown subtraction method\footnote{This dependence on the subtraction scheme can be fixed referring it to the cosmological constant \cite{Filippov:2024kzj}, but here we do not discuss this problem.}.  Then the effective potential in the leading logarithmic approximation becomes completely definite and obeys the generalised RG-equations, which are nonlinear partial derivative equations. The calculations are verified using the limit in which the renormalisable $\ph^4$-interaction model is recovered. The result of latter are known from the literature. Other cases, even for power-like potentials, can be analysed only qualitatively in the framework of different approximations.  In addition, in this paper we have performed numerical evaluation of the implications of summing all-loop leading logarithmic contributions within cosmological phenomenology and discussed the possibility of using the resulting effective potentials in the estimating cosmological observables.

In this paper, we used the expansion near the critical dimension $p=4+\delta$ to analyse the RG-equations. We also found that when the regularisation is removed (when $\epsilon \rightarrow0$ tends to zero), the functions $f_1$ and $f_2$ summing the leading singularities tend to zero. In this light, the behavior of sub-leading divergences is of interest: is the regularisation is removed for them as well?

This work can be continued in other directions. For instance, we can concentrate on studying  of generalisations of the RG-equations to the case of the $SO(N)$-model on a curved background (the studying of the $SO(N)$-model on a flat background was carried out in \cite{Iakhibbaev:2024fjf}), studying of generalisations of the work presented in \cite{Barvinsky:1993zg, Odintsov:1990mt,Elizalde:1993qh} to the case of non-renormalisable models. It would be very fruitful to discuss consideration of higher corrections by curvature in the effective action for an arbitrary scalar field model. However, more important for the development of the apparatus is the study of the subleading corrections and the scheme dependence of the subtraction procedure in non-renormalisable models.  It would also be interesting to find points of contact between our approach and the method discussed in \cite{Ageeva:2024qie} and to extend this method to the case of curved backgrounds.

\section*{Acknowledgements}
The authors are grateful to D.I. Kazakov for his valuable comments and reading of the manuscript. The authors would also like to thank I.L. Buchbinder and I.L. Shapiro for clarifications and explanations. 

\bibliographystyle{unsrt}
\bibliography{main}
\end{document}